\documentclass[preprint,showpacs,preprintnumbers,amsmath,amssymb]{revtex4}

\usepackage{graphicx}% Include figure files
\usepackage{dcolumn}% Align table columns on decimal point
\usepackage{bm}% bold math

\begin{document}

\preprint{}

\title{Matrix plots of reordered bistochastized transaction flow tables: 
A United States intercounty migration example}

\author{Paul B. Slater}% 
\email{slater@kitp.ucsb.edu}
\affiliation{%
ISBER, University of California, Santa Barbara, CA 93106\\
}%
\date{\today}% It is always \today, today,
             %  but any date may be explicitly specified
\newpage
\newpage
\begin{abstract}
We present a number of variously rearranged 
matrix plots of the $3, 107 \times 3, 107$ 1995-2000 
(asymmetric) 
intercounty migration 
table for the United States, principally in its bistochasticized 
form (all 3,107 row and column sums iteratively proportionally fitted to 
equal 1). In one set of plots, the counties are seriated on the bases of the 
subdominant (left and right) eigenvectors of the bistochastic matrix. In 
another set, 
we use the ordering of counties in the dendrogram generated 
by the associated strong component hierarchical clustering. Interesting, diverse  
features of U. S. intercounty migration emerge--such as a contrast
in centralized, hub-like (cosmopolitan/provincial) 
properties between cosmopolitan ``Sunbelt'' and provincial ``Black Belt'' counties.
The methodologies employed should also be insightful for the many other 
diverse forms
of 
interesting transaction flow-type data--interjournal citations being an
obvious, much-studied  example, where one might expect that the journals {\it Science}, 
{\it Nature} and {\it PNAS} would display "cosmopolitan" characteristics.
\end{abstract}


\maketitle
\section{Introduction}
\begin{figure}
\includegraphics{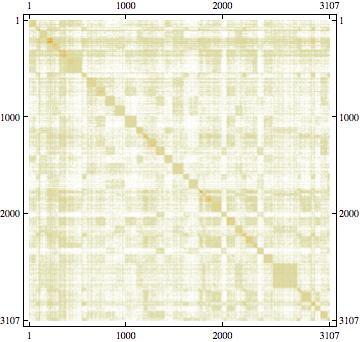}
\caption{\label{fig:matrixplotraw}Unadjusted 1995-2000 intercounty
U. S. migration table. The large square near the end--for
alphabetical reasons--of the
diagonal corresponds to the state with the most (254) counties, Texas, 
while Georgia, with 159 counties, is located near the beginning. County 1000 
is Boyd County, Kentucky and 2000, Dunn County, North Dakota.}
\end{figure}

Based upon the 2000 United States Decennial Census, one can construct a 
square (origin-destination) matrix of 1995-2000 migration flows ($m_{ij}$) 
between
3,107 county-level units of the nation. In Fig.~\ref{fig:matrixplotraw}, 
we show a matrix plot
of this (raw data) 
table. (In the absence of any further relevant information, we set to zero
the diagonal entries--which conceptually 
might correspond either to the number of people who actually moved within 
the county or who simply stayed within it.) 
In the principal, admininstrative  
sorting of the rows/columns of the table, the fifty states are ordered alphabetically, while, secondarily,  
within the states, their constituent counties are ordered also alphabetically. 

We immediately discern a clear clustering along the diagonal in 
Fig.~\ref{fig:matrixplotraw}, indicative of the obvious proposition that migrants have a proclivity to move intrastate-wise, both
for simple distance and state loyalty/ties/allegiance considerations.
However, the alphabetical ordering by states is certainly 
highly fortuitous in
character, and we observe relatively heavy migration far removed from
the diagonal (say for the physically contiguous, but alphabetically 
non-proximate pairs [California, Oregon] and [Texas, Lousiana].)
(Historically, the 
design and layout of counties differs considerably--somewhat 
unfortunately from a geographic-theoretic 
point of view--between states, and we note
that Texas has the most counties, 254, and appears as a large square
far down the diagonal in Fig.~\ref{fig:matrixplotraw}, while 
the state of Georgia, 
with the second most counties, 159, 
is also apparent near the upper left corner.)

\begin{figure}
\includegraphics{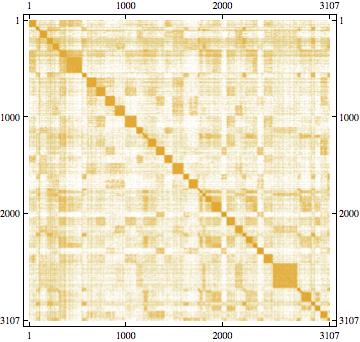}
\caption{\label{fig:matrixplotds}Doubly-stochastic form
of the 1995-2000 intercounty
U. S. migration table}
\end{figure}

Additionally, counties vary widely in population sizes. To control for 
this (marginal) 
effect, one may biproportionally/iteratively adjust the row and 
column sums so that they all converge to be equal (say to 1). In 
Fig.~\ref{fig:matrixplotds}, we show the $3,107 \times 3,107$ intercounty migration 
table after such a double-standardization (bistochastization). 
Clearly, the underlying definition/delimitation of blocks has been heightened 
by this transformation.
The purpose of the scaling is to 
remove overall effects of size (which certainly may be of
interest in themselves), and focus on relative, 
interaction  effects. 
Nevertheless, the {\it cross-product ratios} 
({\it relative odds}), $\frac{m_{ij} m_{kl}}{m_{il} m_{kj}}$, 
measures of association, are left {\it invariant}. 
Additionally, the entries of the
doubly-stochastic table provide 
{\it maximum  entropy} estimates of the original
flows, given the constraints on the row and column sums 
\cite{eriksson,macgill}.
Let us also make the general observations that powers of
bistochastic matrices are also bistochastic, and that 
physicists have been interested in developing conditions
that indicate when a bistochastic matrix is also {\it unistochastic} 
\cite{louck,ZKSS,unistochastic1,dita}. (These latter properties might be of value in the modeling of transaction flows.)
An efficient algorithm--considered as a nonlinear dynamical system--to generate\
 {\it random} bistochastic matrices has
recently been presented \cite{CSBZ} (cf. \cite{griffiths,ZKSS}).

\begin{figure}
\includegraphics{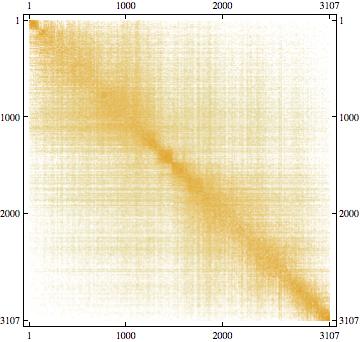}
\caption{\label{fig:subdominant}Doubly-stochastic matrix (Fig.~\ref{fig:matrixplotds}) reordered on the basis of its subdominant left eigenvector. The first
72 counties in the ordering are {\it all} from Georgia (mostly lying in a 
[``Upper Coastal Plain''] band from
the southwest corner of the state [Seminole County] 
to its north central boundary [Franklin, Hart, Elbert and Lincoln Counties]), 
and the last 110, all from the Great Plains states of 
North Dakota (45), South Dakota (50) and (north central) Nebraska (15). County 1000 is
Bucks County, Pennsylvania and 2000, Lubbock County, Texas.}
\end{figure}

\begin{figure}
\includegraphics{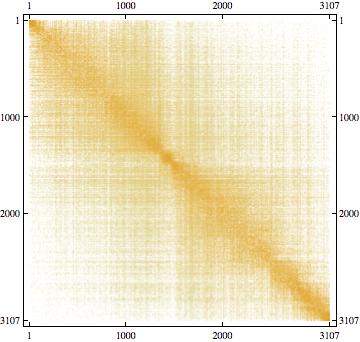}
\caption{\label{fig:subdominantTranspose}Doubly-stochastic matrix (Fig.~\ref{fig:matrixplotds}) reordered on the basis of its subdominant right eigenvector. 
The first 74 counties in the ordering are all from Georgia, and the last 
181, all from North Dakota, South Dakota, Nebraska and Minnesota. County 1000 is Washington County, Louisiana and 2000, 
Adair County, Oklahoma.}
\end{figure}

The dominant left and right eigenvectors (corresponding to the eigenvalue 1) 
of the doubly-standardized table
are simply uniform vectors. The subdominant (left and right) eigenvectors
(corresponding to a {\it real} 
eigenvalue of 0.906253) are of interest 
\cite{meila}. (The correlation between these 
two eigenvectors is high, 0.971197. 
The third largest eigenvalue is real also, 0.868784, while the fourth
is slightly complex in nature, $0.84562 + 0.000906373 i$. The vector 
of 3,107 eigenvalues has length 12.6472.)
We {\it reorder} or {\it seriate} Fig.~\ref{fig:matrixplotds} on the basis of the left (in-migration) 
eigenvector and obtain Fig.~\ref{fig:subdominant}, 
and  on the basis of the right (out-migration) 
eigenvector and obtain Fig.~\ref{fig:subdominantTranspose}.
Now we see diminished clustering far from the diagonal. Further, both
of these figures suggest the division of the nation into basically two
large clusters.

\begin{figure}
\includegraphics{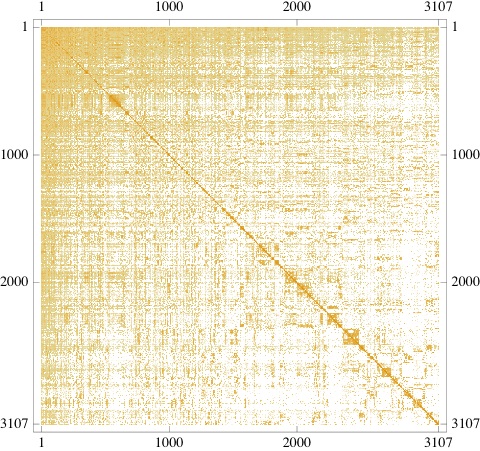}
\caption{\label{fig:ReOrdered1}Doubly-stochastic matrix 
(Fig.~\ref{fig:matrixplotds}) reordered on the basis of its strong component
hierarchical clustering. The first twelve (``cosmopolitan'')
counties in the seriation are
all from the ``Sunbelt'' states of Florida (5 counties,  a 
well-defined cluster of 
four of them
being equivalent to 
the Tampa-St. Petersburg-Clearwater Metropolitan Statistical 
Area), 
Arizona (2), (southern) California (3), Nevada (Las Vegas) (1)
and Texas (Dallas) 
(1). The last 35 (``provincial'') ones--lie 
principally in the ``Black Belt'', stretching  through the Deep South 
states of Mississippi (5),
Alabama (24), Georgia (4) and (Panhandle) Florida (2). County 1000 is Carroll County, 
Indiana and 2000, Warren County, New Jersey.}
\end{figure}

Further, reordering on the basis of the 
(38-page-long, 3,107-county) 
dendrogram (\cite[Supplement]{SlaterDendrogram}) generated by 
the strong component hierarchical clustering (the directed-graph 
analogue of the single-linkage method) \cite{japan,tree,france,science,partial,SEAS,qq,tarjan,tarjan2,ozawa} 
of the bistochastized table, we obtain
Fig.~\ref{fig:ReOrdered1}.  The correlation between the ordering used in this table and the admininstrative ordering used in
Figs. is 0.0373522, and the orderings used in Figs. 3 and 4, respectively, even lower,
0.00401504 and 0.0099957 (Table~\ref{Table}). (The corresponding correlations between the
administrative ordering and that employed in Figs. 3 and 4 are 0.0579257 and 0.0755089. Correlations greater in 
absolute value 
than 0.0353074 are significant at the $95\%$ level, 0.0400655 at the $97.5\%$ level, and 0.0458262 at the $99\%$ level.)
\begin{table} \label{Table}
\begin{tabular}{lllllll}
  & \text{1 $\&$ 2} & 3 & 4 & 5 & 8 & 9 \\
\hline
 \text{1 $\&$ 2} & 1. & 0.0579257 & 0.0755089 & 0.0373522 & -0.00868334 &
   -0.0788444 \\
 3 & 0.0579257 & 1. & 0.140583 & 0.00401504 & 0.00759781 & -0.0202812 \\
 4 & 0.0755089 & 0.140583 & 1. & 0.0099957 & 0.00207526 & -0.000659818 \\
 5 & 0.0373522 & 0.00401504 & 0.0099957 & 1. & 0.0551071 & 0.0206225 \\
 8 & -0.00868334 & 0.00759781 & 0.00207526 & 0.0551071 & 1. & 0.0467724
   \\
 9 & -0.0788444 & -0.0202812 & -0.000659818 & 0.0206225 & 0.0467724 & 1. \\
\hline
\end{tabular}
\caption{Correlations between the orderings of counties 
used in the several 
numbered corresponding figures. 
Correlations greater than 0.0676788 in absolute value 
are significant at the $99.99\%$ 
level, those greater than 0.0458262  at the 
$99\%$ level, and 0.0353074 at the $95\%$ significance level.}
\end{table}

The dominant feature of Fig.~\ref{fig:ReOrdered1} is that 
the counties now listed at the beginning 
in the reordering--and, in general, the last to be absorbed in the
agglomerative clustering process--are ``cosmopolitan'' or 
``hub-like''. They tend to receive and send migrants across the nation, while those nearer to the end in 
the reordering tend to be more provincial or limited
in their breadth of interactions \cite{france}.
(A prototypical example of a hub-like internal migration area is Paris 
\cite{france,winchester}. In analytically parallel studies of interjournal citations 
\cite{science,rosvall,bollen}, one might anticipate that the broad journals, {\it Science}, {\it Nature} and the {\it Proceedings of the National Academy of Sciences} might fulfill analogous roles.)
 
\begin{figure}
\includegraphics{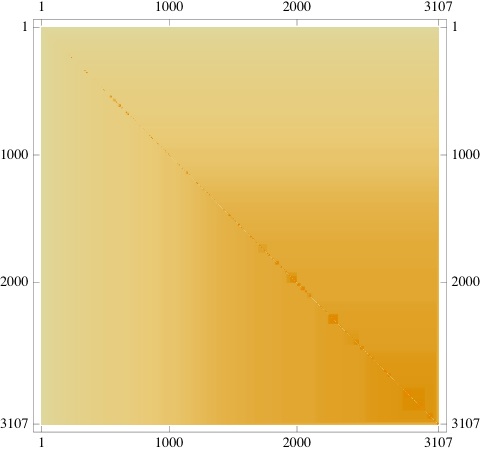}
\caption{\label{fig:ultrametricFitReordered}Ultrametric (strong component 
hierarchical 
clustering) fit to the doubly-stochastic matrix Fig.~\ref{fig:ReOrdered1}. 
The fits tend to be higher in the lower right-hand corner, corresponding
to the more ``provincial'' (including ``Black Belt'') counties.}
\end{figure}

\begin{figure}
\includegraphics{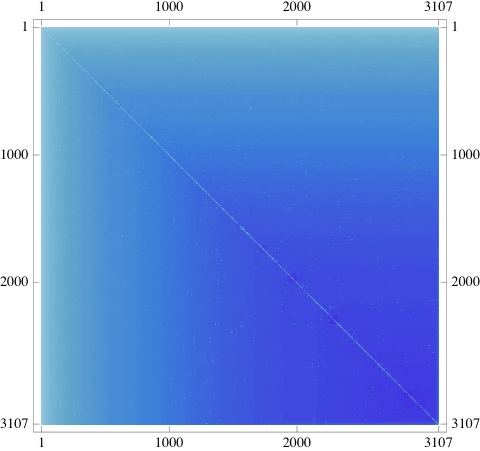}
\caption{\label{fig:residuals}Residuals (predominantly negative) of the 
ultrametric fit (Fig.~\ref{fig:ultrametricFitReordered}) to the doubly-
stochastic matrix (Fig.~\ref{fig:ReOrdered1}). The residuals are most
negative in the lower right-hand corner, where the fits 
provided by the strong component hierarchical clustering were highest.}
\end{figure}

\begin{figure}
\includegraphics{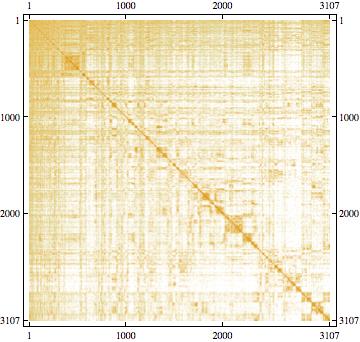}
\caption{\label{fig:direct}
Doubly-stochastic matrix
(Fig.~\ref{fig:matrixplotds}) reordered using the 
hierarchical clustering generated by the DirectAgglomerate
command of Mathematica--the only option in
the Mathematica hierarchical clustering package that seemed 
computationally feasible. The first thirteen counties in the reordering
are from Florida (10), Hawaii (1--Kalawao, the smallest U. S. county) 
and Texas (2), 
while the last twenty-one are 
from Alabama (6), Georgia (10) and Florida (5). County 1000 is Rusk County, Wisconsin and 2000, Knott County, Kentucky.}
\end{figure}

\begin{figure}
\includegraphics{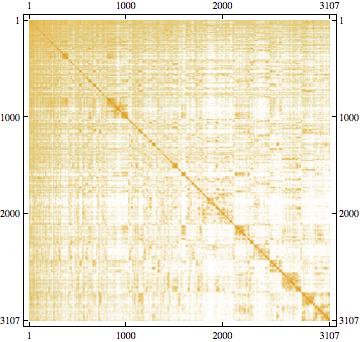}
\caption{\label{fig:directtranspose}
Doubly-stochastic matrix
(Fig.~\ref{fig:matrixplotds}) reordered using the
hierarchical clustering generated by the DirectAgglomerate
command of Mathematica applied to the {\it transpose}. 
The five counties of Hawaii are clustered near the beginning. 
The last thirty-seven counties belong to either Alabama or 
Mississippi. County 1000 is Sciotto County, Ohio and 2000, Polk County, Nebraska.}
\end{figure}

The ultrametric fit to this reordered bistochastized table 
provided by the strong component hierarchical clustering 
\cite{japan,tree,france,science,partial,SEAS,qq,tarjan,tarjan2,ozawa} 
is given in Fig.~\ref{fig:ultrametricFitReordered}, and the residuals 
(predominantly negative) from the fit in Fig.~\ref{fig:residuals}. 
(These latter two figures, both in their  own ways, further 
reflect this cosmopolitan-provincial dichotomy between the 
U. S. counties.)
In Fig.~\ref{fig:direct} we display the bistochastic form of the
1995-2000 U. S. intercounty migration table now reordered on the basis of
the hierarchical clustering generated by application of the 
DirectAgglomerate command of Mathematica. (We inputted our asymmetric values--converted to dissimilarity measures--even though the command assumes a symmetric input. We also applied the same command to the {\it transpose} of 
the dissimilarity matrix, and obtained somewhat differing results 
[Fig.~\ref{fig:directtranspose}].) The correlation
between the orderings in Fig. 8 and Fig. 9 is 0.0467724, and that of the ordering
in Fig. 5 with those in Figs. 8 and 9, 0.0551071 and 0.0206225, respectively.
(With the administrative ordering used in Figs. 1 and 2, the correlations with Figs. 8 and 9 are 
negative, -0.00868334 and [negatively significant] -0.078844, respectively.)

Previously \cite{county,qq}, we have studied (without the aid of 
more recently-developed matrix plots) bistochastized forms of the 1965-70 U. S. intercounty
migration table with strong component hierarchical clustering 
\cite{japan,tree,france,science,partial,SEAS,qq,tarjan,tarjan2,ozawa},
{\it both} with zero and non-zero (corresponding to intracounty 
movements) diagonal entries. Counties with large physical areas tend
to absorb more of their own migrants, and thus exhibit larger
diagonal bistochasticized entries and smaller off-diagonal entries,
making them link at weaker levels in the dendrogram generated. 
Journals with high self-citations would be expected to behave analogously
in journal citation-matrix analyses \cite{science,rosvall,loet}. 
In the application of our two-stage bistochastization and strong component
hierarchical clustering procedure to the 1967-75 interjournal citations between 
twenty-two mathematical journals, the {\it Proceedings of the American
Mathematical Society} was found to function in a 
particularly broad, cosmopolitan 
manner \cite{science}.


\begin{acknowledgments}
I would like to express appreciation to the Kavli Institute for Theoretical Physics (KITP) for technical support.
\end{acknowledgments}

\bibliography{Seriation2009}% Produces the bibliography via BibTeX.

\end{document}